\documentclass[useAMS]{mn2e}
\usepackage{graphicx}
\def\ltsima{$\; \buildrel < \over \sim \;$}
\def\gtsima{$\; \buildrel > \over \sim \;$}
\def\lsim{\lower.5ex\hbox{\ltsima}}
\def\gsim{\lower.5ex\hbox{\gtsima}}
 
\title[The BSS population in NGC~6229] {The BSS population in
  NGC~6229\thanks{Based on observations with the NASA/ESA {\it HST}
    (Prop. 11975), obtained at the Space Telescope Science Institute,
    which is operated by AURA, Inc., under NASA contract
    NAS5-26555. Also based on observations with MegaPrime/MegaCam, a
    joint project of CFHT and CEA/DAPNIA, at the Canada-France-Hawaii
    Telescope (CFHT), which is operated by the National Research
    Council (NRC) of Canada, the Institute National des Sciences de
    l'Univers of the Centre National de la Recherche Scientifique of
    France, and the University of Hawaii.}}

\author[N. Sanna et al.]
 {N.~Sanna$^{1,2}$\thanks{E-mail: nicoletta.sanna2@unibo.it}, E.~Dalessandro$^1$, B.~Lanzoni$^1$, F.R.~Ferraro$^1$, G.~Beccari$^3$, R.T. Rood$^2$ \\
  $^1$Dipartimento di Astronomia, Universit\`a degli Studi di Bologna, via Ranzani 1, I--40127 Bologna, Italy\\
  $^2$Astronomy Department, University of Virginia, P.O. Box 400325, Charlottesville, VA, 22904\\
  $^3$European Southern Obseervatory, Karl-Schwarzschild-Str. 2, D--85748 Garching bei Munchen, Germany} 

\pagerange{\pageref{firstpage}--\pageref{lastpage}}

\begin{document} 

\label{firstpage}

\maketitle

\begin{abstract}
We have used a combination of high-resolution \textit{Hubble Space
  Telescope} WFPC2 and wide-field ground-based observations in
ultraviolet and optical bands to study the blue straggler star (BSS)
population of the outer-halo globular cluster NGC 6229, over its
entire radial extent. A total of 64 bright BSS (with $m_{255}\le
21.30$, corresponding to $m_{555} \le 20.75$) has been identified.
The BSS projected radial distribution is found to be bimodal, with a
high central peak, a well defined minimum at intermediate radii ($r
\sim 40''$), and an upturn in the outskirts.  From detailed star
counts even in the very inner region, we compute the centre of gravity
of the cluster and the most accurate and extended radial density
profile ever published for this system. The profile is reasonably well
reproduced by a standard King model with an extended core ($r_c \simeq
9.5''$) and a modest value of the concentration parameter ($c \simeq
1.49$). However, a deviation from the model is noted in the most
external region of the cluster (at $r>250''$ from the centre). This
feature needs to be further investigated in order to assess the
possible presence of a tidal tail in this cluster.
\end{abstract} 


\section{INTRODUCTION}

In the optical colour-magnitude diagram (CMD) of globular clusters
(GCs) the so called blue stragglers stars (BSS) are bluer and brighter
than the main sequence (MS) objects, appearing younger and more
massive than the normal cluster stars (as also confirmed by direct
mass measurements, e.g. Shara et al. 1997).  Indeed the BSS mass
distribution in 47~Tucanae (47~Tuc) and in NGC~1904, derived from the
star position in the CMD, has been found to peak between 1.1 and 1.2
$M_{\odot}$ (see Ferraro et al. 2006a, Lanzoni et al. 2007a).
This suggests that BSS increase their initial mass during their evolution.
Possible explanations involve the secular (or
induced) evolution of primordial binary systems and collisions between
single and/or binary stars (hereafter PB-BSS and COL-BSS,
respectively; see McCrea 1964, e.g. Bailyn 1995, Sills et al. 1997, and
references therein).

Promising chemical and photometric signatures of different BSS
formation mechanisms have been recently found.  In 47~Tuc Ferraro et
al. (2006a) discovered a subpopulation of BSS showing a significant
depletion of carbon and oxygen, possibly due to a formation through
mass transfer activity in PB-BSS. In M30 two distinct sequences of BSS
have been discovered by Ferraro et al. (2009). The authors argued that
they are populated by BSS originated by the two formation channels,
both triggered by the collapse of the cluster core a few Gyr ago.

In many GCs (M3, 47~Tuc, NGC~6752, M5, M55, NGC~6388, M53, M2, all
references in Beccari et al. 2008; Dalessandro et al. 2008a,b, 2009)
the projected radial distribution of BSS has been found to be bimodal:
highly peaked in the centre, with a clear-cut dip at intermediate
radii, and with an upturn in the external regions.  Dynamical
simulations (Mapelli et al. 2004, 2006; Lanzoni et al. 2007a,b)
suggest that the observed central peak is due to COL-BSS formed in the
core and/or PB-BSS sunk into the centre because of dynamical friction,
while the external rising branch is made of PB-BSS evolving in
isolation in the cluster outskirts. NGC~1904 (Lanzoni et al. 2007a) and M75
(Contreras Ramos et al. 2012)
do not show any bimodality, but BSS appear more segregated in the
central regions than the reference cluster stars. In only three cases
($\omega$~Centauri, Ferraro et al. 2006b; NGC~2419, Dalessandro et
al. 2008b; Pal~14, Beccari et al. 2011), the radial distribution of
BSS is indistinguishable from that of the other stars, suggesting that
these clusters are not significantly mass-segregated yet.

This paper is part of a series devoted to study the UV bright
populations (including horizontal branch -HB, BSS and post-asymptotic
giant branch stars), in old stellar clusters, by means of an extensive
survey conducted with the Wide Field Planetary Camera 2 (WFPC2) on
board the \textit{Hubble Space Telescope} (HST).  More than 30
Galactic GCs have been observed within this project. Here we present a
detailed analysis of the BSS population in NGC~6229. This is one of
the most remote GC associated to outer halo of the Galaxy, located at
$\sim31$ kpc from the Galactic centre (Ferraro et al. 1999). It
probably is a Magellanic Stream member (Palma et al. 2002) and due to
its location, it is poorly studied.  The most recent papers are
focused on HB stars (Borissova et al. 1997, 1999, Catelan et
al. 1998), BSS (Borissova et al. 1999), and variable stars (Borissova
et al. 2001). However, these papers are based on ground-based
observations only, while here we present a study that combines
high-resolution observations from space, thus to properly explore, for
the very first time, the stellar populations in the cluster core, and
wide-field observations from the ground to study the entire cluster
extension.

The paper is organized as follows. Section~2 describes the data set
and the photometric and astrometric analysis. In Section~3 we
determine the centre of gravity and present the CMDs.  The radial
density profile of the cluster is discussed in Section~4. In Section~5
we discuss the BSS properties and present our conclusions. The summary
of the paper is presented in Section~6.

\section{OBSERVATIONS AND DATA ANALYSIS}

In order both to resolve the stars in the crowded central regions, and
to cover the entire extension of the cluster, we used a combination of
high-resolution and wide-field data, similarly to what already done in
our previous studies (e.g. Dalessandro et al. 2009 and references
therein).  The \textit{high-resolution set} consists of a series of
images collected with the HST/WFPC2 at various wavelengths, ranging
from the UV to the optical bands. These images (Prop. 11975,
P.I. Ferraro) were obtained through the UV filter $F255W$ with total
exposure times $t_{\rm exp}= 3600$\,s, and through the optical filters
$F336W$ and $F555W$ with exposure times $t_{\rm exp}= 3000$\,s and
$t_{\rm exp}=245$\,s, respectively.  The centre of the cluster is
located in the Planetary Camera (PC, pixel scale $\sim 0.05''$
pixel$^{-1}$).  The \textit{wide-field set} is composed of
  data obtained with the MegaCam on the Canada-France-Hawaii Telescope
  (CFHT).  A series of 11 images taken through the $g'$ and $r'$
filters with total exposure times $t_{\rm exp}=450$\,s and $t_{\rm
  exp}=540$\,s, respectively, was retrieved from the Canadian
Astronomy Data Centre. MegaCam consists of 36 CCDs of $2048 \times
4612$ pixels each, with a pixel scale of $\sim 0.187''$ pixel$^{-1}$.
The cluster centre is located in chip $\#$23.  The one squared degree
field of view allowed a complete sampling of the cluster.

All the single WFPC2 images were reduced using the DAOPHOTIV package
(Stetson 1987).  In order to model the point spread function (PSF), we
selected bright and almost isolated stars in each frame. We excluded
stars lying in the central regions of the PC for the $F336W$ and
$F555W$ exposures, because of crowding problems. Typically $\sim 200$
$``$bona fide stars$"$ have been used to model the PSF in the $F336W$
and $F555W$, and $\sim 80$ in the $F255W$. With the obtained PSF
models we performed a first PSF-fitting on each single image by using
ALLSTAR. Since our analysis is particularly devoted to hot stars, as
master frame for the following reduction we used the stacked image
obtained combining the $F336W$ and the $F255W$ exposures, using
{\tt{DAOMATCH/DAOMASTER/MONTAGE}} packages (Stetson 1994). Stars found
in the master reference frame have been force-fitted to each single
frame by using the ALLFRAME package (Stetson 1994).  At this point we
selected again the PSF stars from the ALLFRAME catalogue and we
repeated the whole procedure.  In this way we have minimized the risk
of having spurious detections between stars used for the PSF modelling
and the advantage of choosing $``$bona fide stars$"$ on the basis of
more reliable magnitude and improved centroid determination.

The reduction procedure for the MegaCam images is the same as for the
WFPC2. In this case, however, the master list for each chip is made by
stars identified in at least two out of 11 exposures. This choice
allowed us to get full advantage of the dithering strategy adopted for
these observations.

All the catalogues were put on the absolute astrometric system using a
large number of stars in common with the Sloan Digital Sky Survey
(SDSS) catalogue. As a first step we obtained the astrometric solution
for each of the 36 chips of MegaCam by using the procedure described
in Ferraro et al. (2001, 2003) and the cross-correlation tool
CataXcorr (Montegriffo, private communication).  All the stars in
common with the HST field were then used as secondary astrometric
standards in order to put all the catalogues in the same astrometric
system. Several hundred astrometric standards have been used in each
step, allowing a very precise astrometry for each catalogue. At the
end of the procedure the estimated error in the absolute positions,
both in right ascension ($\alpha$) and declination ($\delta$), is of
about $0.2''$.

The instrumental magnitudes of the high-resolution set were corrected
for charge transfer efficiency by using the equations of Dolphin
(2009). All the WFPC2 magnitudes ($m_{255}$, $m_{336}$ and $m_{555}$)
were calibrated to the VEGAMAG system by using the prescription by
Holtzman et al. (1995) and the zero points from the WFPC2 data
handbook.\footnote{http://documents.stsci.edu/hst/wfpc2/documents/handbooks/dhb}
In order to calibrate the MegaCam magnitudes, we first transformed the
$g'$ and $r'$ instrumental magnitudes to the SDSS system ($g$ and $r$)
by using the stars in common with the Sloan public catalogues. We then
converted the $g$ magnitudes in the $m_{555}$ VEGAMAG by means of the
following colour equation: $m_{555}=g-0.501*(g-r)+0.256$, obtained by
using the stars in common between the two catalogues.

In order to exclude extra-galactic sources from our analysis, we
retrieved a catalogue of these objects from the NASA EXTRAGALACTIC
DATABASE\footnote{http://ned.ipac.caltech.edu/} and we matched it with
our catalogues.  We found a total of $\sim 2880$ objects in common
(typically galaxies and quasars) which were excluded from the
following analysis.

\section{CENTRE OF GRAVITY AND CMDs}
Thanks to the high-resolution and quality of the WFPC2 images, we
determined the centre of gravity $C_{grav}$ of NGC~6229 from resolved
stars, by following the iterative procedure described in Montegriffo
et al. (1995; see also Ferraro et al. 2003, 2004; Lanzoni et
al. 2007c; Dalessandro et al. 2009). We used the centre quoted by
Harris (1996, 2010 version) as the first guess, and the average value
of the $\alpha$ and $\delta$ positions of all the stars contained
within circles of different radii (from $8''$ to $13''$, stepped by
$1''$) as the new centre, until convergence was reached.  In order to
avoid possible spurious effects due to incompleteness of the
catalogue, we considered three samples with different limiting
magnitudes ($m_{555}= 21.0, 21.3, 21.5$). The obtained values
  agreed within $\sim 0.3''$, and their average was therefore assumed
  as $C_{grav}$: $\alpha (J2000.0)=16^{\rm h} 46^{\rm m} 58.74^{\rm s}
  \pm 0.28''$, $\delta (J2000.0)=47^{\circ} 31' 39.53'' \pm
  0.13''$. This new determination is located $\sim 0.6''$ South-West
  ($\Delta\alpha\simeq-0.5''$, $\Delta\delta\simeq-0.4''$) from the
  Harris centre.

We then used the computed cluster centre to divide the entire data set
in two samples which most suitably describe different regions of the
system.  The \textit{HST sample} is composed of 17307 stars measured
in the WFPC2 observations and located at a distance $r<90''$ from
$C_{grav}$ (see Figure~1).  While this region is not entirely sampled
by the peculiar WFPC2 field of view (FOV), according to previous work
(see Sabbi et al. 2004) and in order to avoid severe incompleteness
effects, we conservatively preferred to not complement this sample
with low-resolution ground-based observations.  The \textit{External
  sample} includes all the stars measured in the MegaCam FOV at
$r>90''$, and it counts 28210 stars (see Figure~2).

\begin{figure}
\includegraphics[scale=0.43]{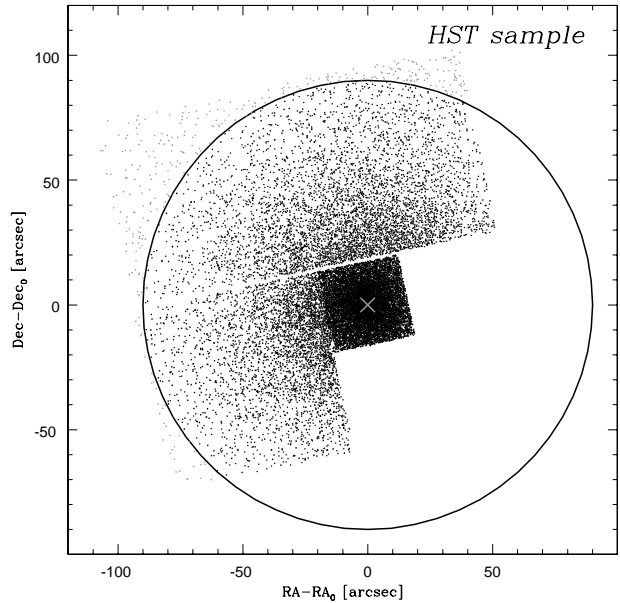}
\caption{Map of the \textit{HST sample}. Position of the stars
  measured in the WFPC2 FOV plotted with respect to the centre of
  gravity (large cross). The circle at $r=90''$ marks the adopted edge
  of the \textit{HST sample}.}
\end{figure}

\begin{figure}
\includegraphics[scale=0.43]{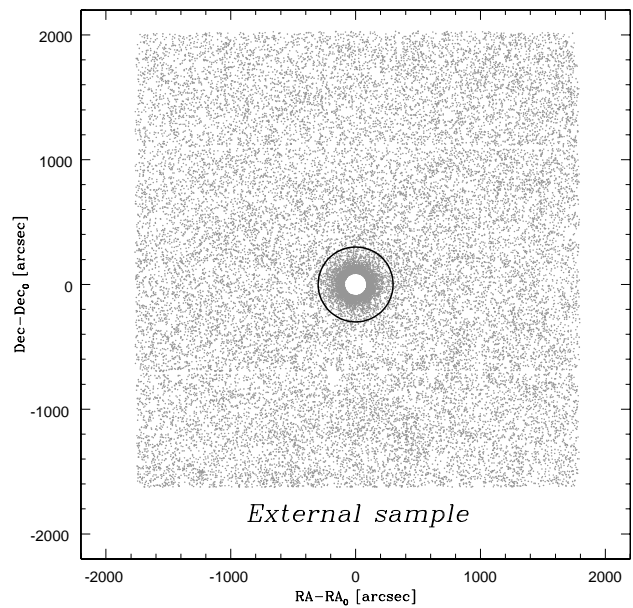}
\caption{Map of the \textit{External sample}. Only stars observed with
  MegaCam at $r>90''$ have been included in this sample.  The circle
  marks the nominal tidal radius of the cluster ($r=300''$) as set by
  the best-fit King model shown in Figure~6.}
\end{figure}

The CMDs of these samples are shown in Figures~3, 4, 5.  As apparent
from Figure~3, the evolutionary sequences display the morphology
typical of the UV plane (see Ferraro et al. 1997, 2003), with the HB
crossing diagonally the diagram, the red giant branch (RGB) stars
extending to colours $(m_{255}-m_{336}) \sim 6$ and always being below
$m_{255} \sim 21.5$, and the BSS population defining a vertical
sequence at $(m_{255}-m_{336}) \sim 0.5$, which merges without any
discontinuity into the turn-off of the MS at $m_{255} \sim 22.5$.  The
same stars are plotted for comparison in the
($m_{555},m_{336}-m_{555}$) plane in the left panel of Figure~4.  The
comparison of these two figures shows that at the optical wavelengths
the cool objects (as the RGB stars) are much brighter than the hot
objects (like BSS and HB stars), which instead dominate the bright
portion of the CMD at UV wavelengths.  For this reason, we decided to
select the hot stars of the \textit{HST sample} in the UV plane and
the cool objects in the optical CMDs.

The HB of NGC~6229 has been subject of a number of studies.  As
suggested by Borissova et al. (1997) and Catelan et al. (1998) the HB
morphology is characterized by a bimodal distribution in $(B-V)$
colour and shows at least one gap on the blue HB at $V\sim18.4$.
NGC~6229 is the only outer-halo cluster and one of the four known
cases (the others being NGC~2808, NGC~6388 and NGC~6441) where these
two anomalies are simultaneously present.  Moreover, Borissova et
al. (1999) identified 9 extreme HB stars in this cluster.  The
availability of high-resolution CMDs, both in the UV and in the
optical planes, clearly is a major advantage for a more detailed study
of the HB morphology in NGC~6229. While this will be discussed in a
forthcoming paper, here we use the HB stars only as reference
population (see Section 5.2). The right panel of Figure~4 shows the
($m_{555},m_{555}-m_{r}$) CMD for the \textit{External sample}. As can
be seen, the cluster population is well defined and largely dominant
over the field stars out to $r\sim300''$.  Indeed the MS of NGC~6229
can be recognized even beyond this distance (see Figure~5), while the
Galactic field contamination becomes clearly dominant in the most
external regions sampled by our data set.

\begin{figure}
\includegraphics[scale=0.43]{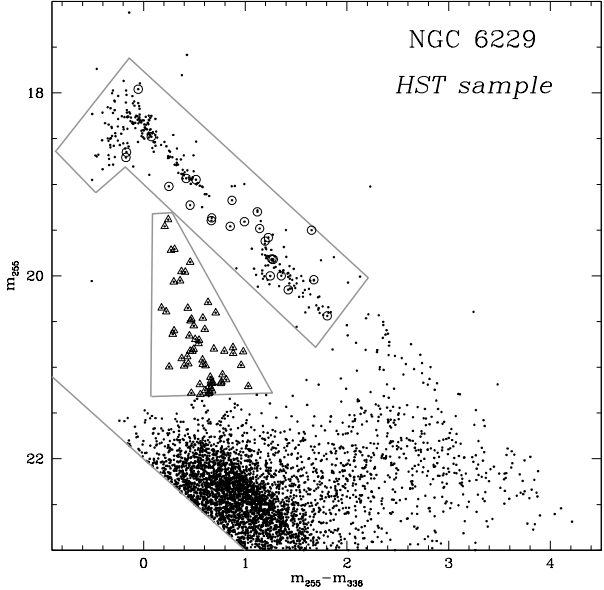}
\caption{UV CMD of the \textit{HST sample}. The adopted boxes used for
  the selection of the BSS (empty triangles) and the HB populations
  are shown. Known RR Lyrae stars (from Borissova et al. 2001) are
  marked with empty circles.}
\end{figure}

\begin{figure}
\includegraphics[scale=0.43]{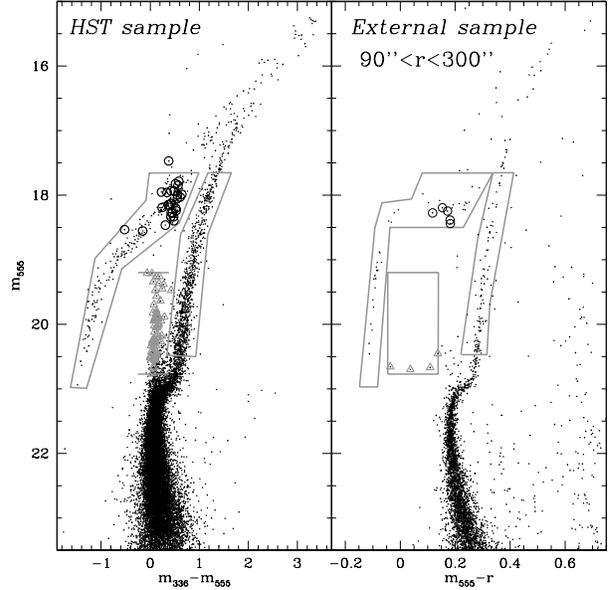}
\caption{Optical CMDs of the \textit{HST} and the \textit{External}
  samples. The adopted selection boxes for BSS, RGB and HB stars are
  shown.  BSS are plotted as empty triangles (those of the \textit{HST
    sample} have been selected in the UV CMD shown in Figure~3).  The
  $m_{555}$ magnitude range for the BSS selection is also shown in the
  left panel by two horizontal grey segments.  As in Figure~3, known
  RR Lyrae are marked with open circles.}
\end{figure}

\begin{figure}
\includegraphics[scale=0.43]{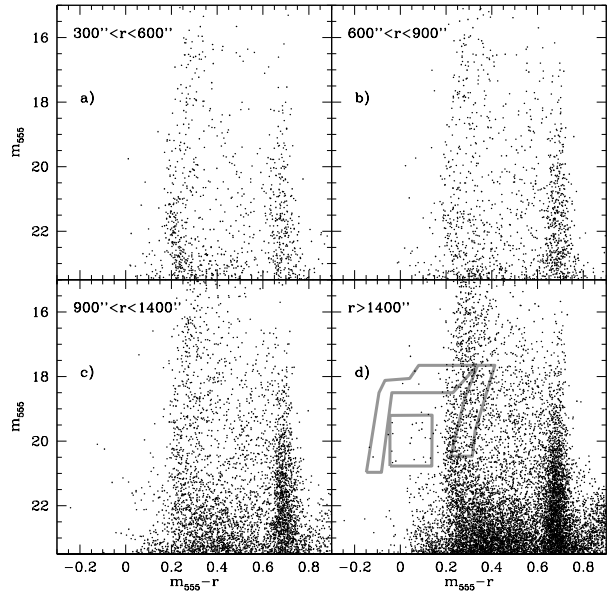}
\caption{CMDs of the \textit{External sample} at different radii. The
  cluster MS is visible (at $21<m_{555}<23$) even beyond the nominal
  cluster tidal radius ($r_t=300''$) obtained by the best-fit King
  model (see Panel a). Panel d) shows the region used to determine the
  field star contamination; the adopted selection boxes for BSS, RGB
  and HB stars are also marked.}
\end{figure}

\section{RADIAL DENSITY PROFILE}
We determined the projected density profile of NGC~6229 by measuring
the star counts over the entire cluster extension, from $C_{grav}$ to
$r \sim 2200''$.  From the \textit{HST sample} we selected fiducial
RGB, subgiant branch and bright MS stars with $19.5<m_{555}<21.5$. The
fiducial stars are those within three times the typical photometric
error from the ridge mean line of each evolutionary sequence.  The
same criterium was adopted to select stars from the \textit{External
  sample}. In this case we considered the magnitude interval
$21.0<m_{555}<23.0$, thus to limit the strong contamination from field
stars in the most external regions, which would be particularly severe
along the (scarcely populated) RGB.  We divided the entire FOV in 18
annuli centred on $C_{grav}$ and each annulus was divided into two or
four subsectors, where we computed the ratio between the number of
stars and the subsector area. The stellar density of each annulus is
then obtained from the average of the corresponding subsector
densities and the errors from the squared root of their variance. The
incomplete area coverage affecting some annulus has been taken into
account in this procedure.  In order to join the star counts in the
inner and outer regions, we determined the density of the
\textit{External sample} in the annulus $60''<r<90''$ using the same
magnitude range adopted for the \textit{HST sample}
($19.5<m_{555}<21.5$) and we used this point to normalize the two
distributions.  The resulting radial density profile is shown in
Figure~6.  The last three points ($r>750''$) have been used to
estimate the contribution of the background stars.

The single mass King model that best fits the observed density profile
(see the solid line on figure) has concentration $c\simeq1.49$ and
core radius $r_c\simeq9.5''$.  These values are consistent with those
($c=1.50$, $r_c=7.7''$) quoted by McLaughlin \& van der Marel (2005),
who however fit the surface brightness (instead of the surface
density) profile, only out to $r\sim100''$.  Our fit well reproduces
the central part of the profile and confirms that NGC~6229 has not
experienced the collapse of the core yet, as suggested by Djorgovski
\& King (1986), Trager et al. (1995), Borissova et al. (1997).
However, neither our best-fit model, nor those found in the literature
are able to properly reproduce the external density profile (see the
figure).  In fact, the best-fit King model systematically
underpredicts the number counts for $r>250''$ and a prominent MS
belonging to the cluster is well appreciable beyond the nominal tidal
radius of the model, set at $r\sim300''$ (see Figure~5).  This
discrepancy requires a more detailed investigation, which is out of
the scope of the present paper.

\begin{figure}
\includegraphics[scale=0.43]{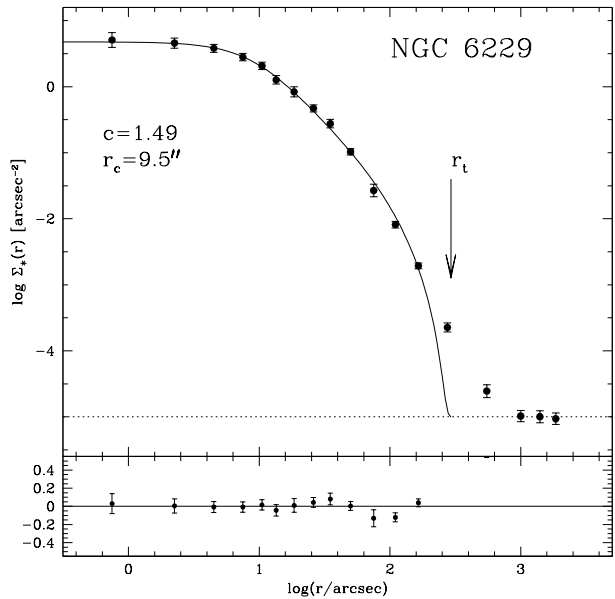}
\caption{Observed surface density profile (filled circles) and
  best-fit King model (solid line). The radial profile is in unit of
  number of stars per square arcsecond. The dotted line indicates the
  adopted level of the background, computed as the average of the
  three outermost points.  The location of the nominal cluster tidal
  radius is marked by the arrow. The lower panel shows the residuals
  between the observations and the fitted profile for $r\leq200''$.}
\end{figure}

\section{THE BSS POPULATION}
\subsection{\textit{The BSS selection}}
For the \textit{HST sample} we selected the BSS population in the
($m_{255},m_{255}-m_{336}$) diagram. The adopted selection box is
shown in Figure~3.  In order to avoid the contamination from  MS
turn-off stars, we limit the sample to $m_{255}<21.3$. Within this
threshold, we identify 60 BSS in the \textit{HST sample}, which are
also shown in the left panel of Figure~4 and turn out to span the
magnitude range $19.2<m_{555}<20.75$.  Because of the quality of the
optical CMD, the BSS selection in the \textit{External sample} is
quite straightforward.  We used the same magnitude limits inferred in
the optical plane by stars selected in the UV, while the extension in
colour has been set by the four candidate BSS visible in this diagram
at $-0.05<(m_{555}-r)<0.15$. The selection box is shown in grey in the
figure.  Interestingly enough the most external BSS lies at $\sim
130''$ from the centre, i.e. well within the cluster tidal radius.
Thus the final sample of BSS counts 64 objects in the surveyed region
of the cluster.

A preliminary study of the BSS population at $r>32''$ has been
presented by Borissova et al. (1999), using optical ground-based data.
It is not possible to compare that sample with our much higher
resolution HST data covering the central and most crowded region of
the cluster.  However, when comparing the ground-based data in the
outskirts of the cluster, we find that the number (4) of BSS at
$r>90''$ in the \textit{External sample} is in good agreement with the
number (5) quoted by Borissova et al. (1999).

\subsection{\textit{The reference populations}}
As already discussed in other papers (see e. g. Ferraro et al. 1995,
1997) in order to study the BSS properties, we need to select also a
reference population representative of the normal cluster stars. To
this end, we consider both the HB and the RGB populations.  Since the
HB of NGC~6229 is bimodal (Borissova et al.\ 1997; Catelan et
al.\ 1998) and it is populated by stars cooler and hotter than the
instability strip, we decided to select HB stars by using both the UV
and the optical planes, when possible.  In fact the ($m_{255}$,
$m_{255}-m_{336}$) guarantees a solid selection of the blue HB stars,
while it is strongly incomplete for stars with
$(m_{255}-m_{336})>2.5$, which are instead among the brightest objects
in the optical plane. We first selected HB stars in the UV plane by
using the selection box shown in Figure~3.  Thanks to the adopted
photometric reduction method, all the stars selected in this way are
also identified in the ($m_{555}$, $m_{336}-m_{555}$) plane and they
lie at $(m_{336}-m_{555})<0.5$.  We then built a selection box in the
optical CMD including both the stars identified in the UV and the HB
stars redder than $(m_{336}-m_{555})=0.5$ (see left panel of
Figure~4). This selection allows us to identify all the HB stars and
to obtain the magnitude limits for the selection in the outermost
regions of the cluster.  In fact, since no UV filters are available
for the \textit{External sample}, we selected HB stars in the
($m_{555},m_{555}-r$) CMD by using the same magnitude limits
($17.5<m_{555}<21.0$) used in the optical CMD of the \textit{HST
  sample}.  All the 30 known RR Lyrae stars at $r>15''$ (Borissova et
al.\ 2001) located in our FOV have been included in the HB selection.
Considering the entire FOV, from $C_{grav}$ to $r\sim300''$, we
identify 339 HB stars, 299 in the \textit{HST sample} and 40 in the
\textit{External sample}.  To select the RGB stars we used the optical
CMDs, where these objects are bright and the branch well defined.  In
order to reduce the contamination from subgiant and asymptotic giant
branch stars, we limit our selection to $17.5<m_{555}<20.5$.  We
identify 1027 RGB stars, 930 in the \textit{HST} field and 97 in the
\textit{External} field.  The boxes adopted to identify the reference
populations are shown in both panels of Figure~4.

\subsection{\textit{The BSS radial distribution}}
Figure~7 shows the cumulative radial distribution for the three
selected populations (namely BSS, HB and RGB).  As evident, BSS are
more centrally concentrated than the others. The Kolmogorov-Smirnov
(K-S) test gives a probability of $\sim7$x$10^{-9}$ and
$\sim6$x$10^{-9}$ that BSS are extracted from the same parent
populations of RGB and HB stars, respectively.

\begin{figure}
\includegraphics[scale=0.43]{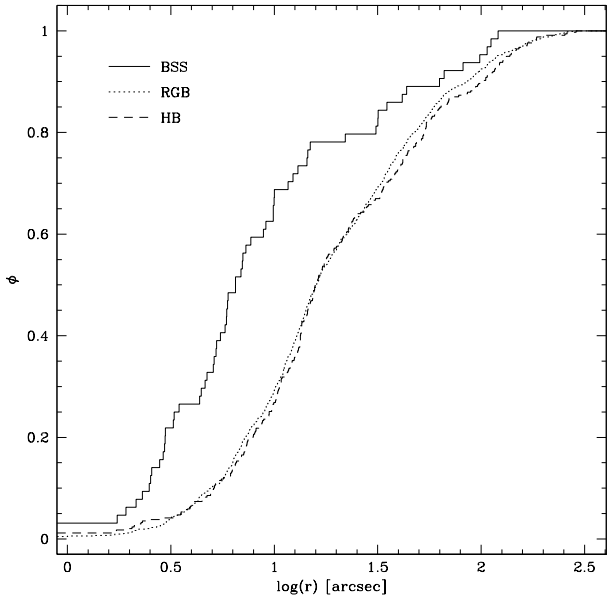}
\caption{Cumulative radial distribution of BSS (solid line), HB (dashed line) 
and RGB (dotted line) stars as a function of the projected distance from
$C_{grav}$.}
\end{figure}

For a more quantitative analysis, we computed the population ratios
$N_{\rm {BSS}}/N_{\rm {HB}}$, $N_{\rm {BSS}}/N_{\rm {RGB}}$ and
$N_{\rm {HB}}/N_{\rm {RGB}}$ in seven concentric annuli centred in
$C_{grav}$.  In order ensure that our results are not affected by
severe field contamination, we carefully evaluated the expected number
of field stars in each selection box. To this end we exploit the huge
FOV covered by the MegaCam catalogue, that allows us to statistically
quantify the contamination of the field stars well beyond $r=300''$.
In particular we used the CMD obtained for $r>1400''$, where field
stars define two vertical sequences roughly located at
$0.2<(m_{555}-r)<0.4$ and $0.5<(m_{555}-r)<0.8$.  By counting the
number of stars lying within the boxes used for the population
selections, we derived the following values for the field star
densities: $\rho_{\rm {BSS}}\sim0.010$ stars arcmin$^{-2}$, $\rho_{\rm
  {HB}}\sim0.027$ stars arcmin$^{-2}$, $\rho_{\rm {RGB}}\sim0.102$
stars arcmin$^{-2}$. This indicates that the BSS and the HB
  samples are not very affected by field stars contamination within
  $300''$. Indeed, after the decontamination from field stars, we
obtain $N_{\rm {BSS}}=64$, $N_{\rm {HB}}=337$, $N_{\rm {RGB}}=1019$,
and a total specific frequency of BSS $F^{\rm {BSS}}_{\rm {HB}}=N_{\rm
  {BSS}}/N_{\rm {HB}}\simeq0.19$.

\begin{table}
\begin{tabular}{|p{1.3cm}||p{1.3cm}||p{1.3cm}||p{1.3cm}|*{5}{c|}|}
\hline
$r''_i$&$r''_e$&$N_{\rm {BSS}}$&$N_{\rm {HB}}$&$N_{\rm {RGB}}$\\
\hline
0&5&21&33&105\\
5&15&29&125&385\\
15&35&4&80&251\\
35&60&3&46&136\\
60&90&3&15&53\\
90&130&4&21&53 (1)\\
130&300&0&19 (2)&44 (7)\\
\hline
\end{tabular}
\caption{Number of BSS, HB and RGB stars counted in the seven
  concentric annuli used to study the BSS radial distribution.  The
  values in parenthesis are the estimated number of contaminating
  field stars in each annulus (see text, Section~4.2).}
\end{table}

The star counts for each annulus are listed in Table 1 and have been
used to compute the population ratios as a function of the radial
distance from the cluster centre (see Figure 8).  The BSS distribution
is clearly bimodal, with a high peak in the centre, a minimum at
$r\sim40''$ and a rising branch in the outer region.  In contrast, the
$N_{\rm {HB}}/N_{\rm {RGB}}$ ratio (bottom panel of the figure) is flat
across the entire extension of the cluster, as expected for normal,
not segregated populations.  The central segregation of BSS is quite
high. In fact, $\sim67\%$ of the entire BSS sample is located within
$r\le10''$, while only $\sim27\%$ of the reference stars is found in
the same region. By assuming the central luminosity density
  quoted by Harris (1996, 2010 version) and the central velocity
  dispersion (6.8 km s$^{-1}$) published by McLaughlin \& van der
  Marel (2005), the resulting radius of avoidance (see e. g. Mapelli
  et al. 2006) would be $r_{av}\simeq75''$. This value is larger than
  the observed position of the minimum of the radial
  distribution. However, we note that a (reasonably larger) velocity
  dispersion of $\sim 9$ km s$^{-1}$ would be sufficient to bring into
  agreement these two quantities. In addition, in such a computation
  we are assuming a King model that underestimates the cluster density
  in the outer regions, and this could also be responsible for at
  least part of the discrepancy. Since a more precise analysis is not
  possible at the moment we can only conclude that the central regions
  of NGC~6229 likely are already relaxed, while its outskirts still
  are not much affected by dynamical friction effects.

\begin{figure}
\includegraphics[scale=0.43]{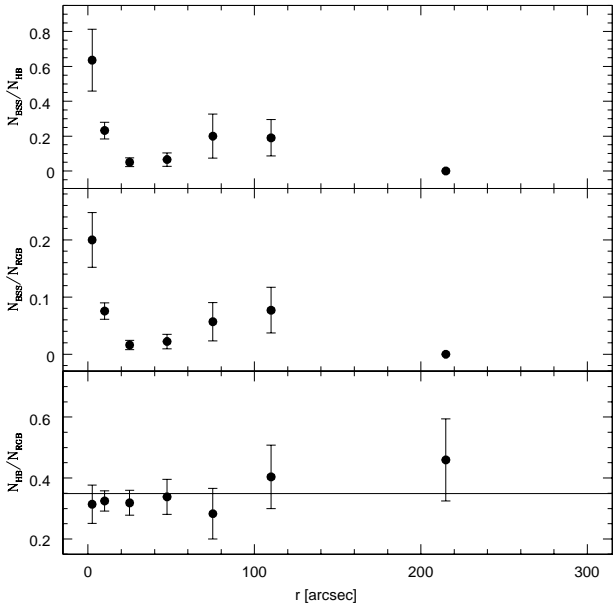}
\caption{Distribution of the population ratios $N_{\rm {BSS}}/N_{\rm
    {HB}}$, $N_{\rm {BSS}}/N_{\rm {RGB}}$, $N_{\rm {HB}}/N_{\rm
    {RGB}}$ (top, middle, bottom, respectively) as a function of the
  radial distance from the cluster centre.}
\end{figure}

\section{SUMMARY}
In this paper we have used a combination of HST UV and optical images
of the cluster centre, and wide-field ground-based optical
observations covering the entire cluster extension to derive the main
structural parameters of the globular cluster NGC~6229 and to study
its BSS population.

From the high-resolution data we derived the cluster centre of gravity
lying at $\alpha (J2000.0) = 16^{\rm h} 46^{\rm m} 58.74^{\rm s}$,
$\delta (J2000.0) = 47^{\circ} 31' 39.53''$, with an uncertainty of
$\sim 0.3''$ in both $\alpha$ and $\delta$. We determined the radial
density profile from star counts, finding that the King model that
best fits the central region (out to $\sim250''$) is characterized by
a core radius $r_c\simeq9.5''$ and a concentration
$c\simeq1.49$. However such a model does not well reproduce the
outermost portion of the cluster profile.

A total of 64 BSS (60 in the \textit{HST sample} and 4 in the
\textit{External sample}) has been identified.  The radial
distribution of BSS with respect to normal cluster stars (HB and RGB)
is bimodal: with a high peak in the centre, a clear-cut dip at
intermediate radii ($\sim40''$), and an upturn in the external
regions. Such a bimodality is quite similar to that found in the
majority of the GCs investigated so far, and it has been interpreted
as the result of the dynamical relaxation of the clusters
(e. g. Beccari et al. 2008; Dalessandro et al. 2009; see also Section 1).

\section*{ACKNOWLEDGMENTS}
This research used the facilities of the Canadian Astronomy Data Centre
operated by the National Research Council of Canada with the support of the 
Canadian Space Agency.
This research is part of the project COSMIC-LAB founded by the
European Raeserch Council (under contract ERC-2010-AdG-267675).
Financial contribution of the Italian Istituto Nazionale di
Astrofisica (INAF, under contract PRIN-INAF 2008) and the Agenzia
Spaziale Italiana (under contract ASI/INAF/I/009/10) is also
acknowledged.  The research leading to these results has received
funding also from the European Community's Seventh Framework Programme
(/FP7/2007-2013/) under grant agreement No 229517.

\label{lastpage}
\end{document}